\documentstyle{l-aa}
\input{epsf}

\def\spose#1{\hbox to 0pt{#1\hss}}
\def\simlt{\mathrel{\spose{\lower 3pt\hbox{$\mathchar"218$}}
     \raise 2.0pt\hbox{$\mathchar"13C$}}}
\def\simgt{\mathrel{\spose{\lower 3pt\hbox{$\mathchar"218$}}
     \raise 2.0pt\hbox{$\mathchar"13E$}}}
\def\mpc {h^{-1} {\rm{Mpc}}}
\def\and  {{\it {et al.} }}
\def\disp{\displaystyle}
\def\gam{{\bf \gamma}}
\def\d{{\rm d}}
\def\mg{\big<}
\def\md{\big>}
\def\vr{{\bf r}}
\def\vx{{\bf x}}
\def\vk{{\bf k}}
\def\ii{{\rm i}}
\def\grad{\nabla}
\def\be{\begin{equation}}
\def\ee{\end{equation}}
\def\ba{\begin{eqnarray}}
\def\ea{\end{eqnarray}}

\input psfig.sty

\begin{document}

   \thesaurus{12 (12.03.1; 12.04.1; 12.07.1; 12.12.1)} 

 \title{The skewness and kurtosis
of the projected density distribution function: 
validity of perturbation theory}

\author{E. Gazta\~naga$^{1,2}$, F. Bernardeau$^3$}

 \offprints{F. Bernardeau; fbernardeau@cea.fr}

 \institute{
$^1$Institut d'Estudis Espacials de Catalunya, Unidad Nixta del CSIC,
Edf. Nexus-104 - c/ Gran Capitan 2-4, 08034 Barcelona \\
$^2$ NASA/Fermilab Astrophysics Center, Fermilab, Batavia, IL 60510 \\
$^3$Service de Physique Th\'eorique, 
C.E. de Saclay, F-91191 Gif-sur-Yvette cedex, France}

\maketitle

\markboth{Skewness of the projected density}{E. Gazta\~naga \& F. Bernardeau}

\begin{abstract}

We study the
 domain of validity of Perturbation Theory (PT), 
by comparing its predictions for the reduced skewness, 
$s_3$, and kurtosis, $s_4$, 
of the projected cosmological density field, with the results
of N-body simulations. We investigate models with 
different linear
power spectra and consider as 
physical applications both 
angular galaxy catalogues and weak lensing surveys.
We first find that the small-angle approximation for the
predicted skewness
provides a good match to the exact  numerical PT results.   
On the other hand, 
results from non-linear simulated catalogues agree well
with PT results on quasi-linear angular scales, which 
correspond to scales larger than about $1 \, deg$ in the 
applications we  have considered.
We also point out that, on smaller scale, the projection effects 
tend to attenuate the effects of the
strong nonlinearities in the angular skewness and kurtosis.


\end{abstract}

\section{Introduction}

One of the principal goal of the large galaxy surveys that should be available
in the near future (SDSS, 2DF, deep surveys with the MEGACAM project, ...) 
is the accurate determination of the
power spectrum of density fluctuations, $P(k)$.
Its interpretation in terms of cosmological models, however, requires an
understanding of the way galaxies trace the underlying
matter distribution. This is a general problem that is bound to become
even more crucial with these
new observational data.

It has been stressed recently (Fry \& Gazta\~naga 1993, 
Gazta\~naga \& Frieman 1994, Fry 1996, Bernardeau 1995, hereafter B95) 
that one way to address this problem
is to consider higher-order correlation functions or higher-order moments
of the local density probability distribution function. This approach
is particularly attractive because, in the 
case of Gaussian initial conditions, 
Perturbation Theory (PT) provides precise quantitative predictions. 
It has now been well established that the $p$-order cumulants
of the local density field $\mg\delta^p\md_c$ are expected
to behave as
\be
\mg\delta^p\md_c=S_p\,\mg\delta^2\md^{p-2}
\ee
on large scales (Fry 1984, Goroff et al. 1984, Bouchet et al. 1992,
Bernardeau 1992). The $S_p$
parameters, which quantify the departure from Gaussian behavior,
depend however on the window function applied to the field.
In Bernardeau (1994) a prescription is given for the PT
calculation of all these
coefficients for a 3D top hat window function. These results
have been subsequently checked in detail and found to be very accurate
when compared to numerical simulations (Baugh, Gazta\~naga \& Efstathiou 1994,
Baugh \& Gazta\~naga 1995). Perturbation results for the 3D Gaussian window
function (Goroff et al. 1986, Juszkiewicz, Bouchet \& Colombi 1993)
for $S_3$ and $S_4$ have also been successfully tested against numerical
simulations. 
Another fruitful direction of investigation is the calculation of 
the high-order correlation functions for the projected density
(B95). This is an interesting domain  
of investigation since angular galaxy catalogs contain more objects 
and volume than 
3D catalogs and therefore allow the determination of many more
parameters of the hierarchy (1).

There is also a new mean of investigation, which is still in an embryonic
state but might reveal extremely fruitful, for the projected density.
The measurement of the gravitational weak shear induced by the 
large scale structures in deep
galaxy catalogs allow in principle to have access to the correlation properties
of the projected mass. The resulting polarization
maps could allow the determination of these correlation functions 
at the level of the two-point function
(Blandford et al. 1991, Miralda-Escud\'e, 1991, Villumsen 1996,
Jain \& Seljak 1997, Kaiser 1996) or even for higher
orders (Bernardeau, van Wearbeke \& Mellier 1997). 

Although the projected density is obtained with completely
different methods in galaxy catalogs or in weak lensing surveys,
the properties of the projected density as yielded by Perturbation Theory
are addressed in very similar ways. These two cases differ
only in the shape of the required selection function, whereas identical
approximations are made in the course of the calculations
for the derivation of the analytic expressions.
We recall here that Perturbation Theory results are valid at large scales,
where the variance is small. For 3D filtering,
investigations in numerical simulations 
have proved, for 3D filtering, that PT results are
valid for scales above about
$10\,h^{-1}$Mpc, where the variance of fluctuations approaches unity.
In the case of the projected density, however,
a given angular scale cannot be straightforwardly associated
with a physical scale, since it corresponds to a
superposition of different scales. A priori 
it is therefore difficult to assess the validity
domain of PT results, even in light of the 3D cases. 
Another concern is the use of the small
angle approximation. This is a mathematical approximation
that allows one to  dramatically simplify
the calculations. So far it has
 not been possible to get closed analytic formulae
without its use. This approximation might be not fully
valid when the smoothing angle is above 1 degree (see B95).

The aim of this paper is therefore to investigate the validity
of both the small angle and the PT approximations. 
Numerical simulations with $N$-body codes are used to check
the validity of PT at small scales, and to give better clues to
the importance of the fully non-linear corrections that may affect
the quantities we are interested in.
The effect of the small-angle approximation is more particularly 
investigated by direct Monte-Carlo integration of the
$s_3$ coefficient. In this case, the problem is not
the validity of the Perturbation Theory approach, but the 
accuracy of the mathematical approximations that were
made in the course of the calculation.

In \S 2, we specify the models that are used
to illustrate those calculations. We also give the expected
values for $s_3$ in those cases.
In \S 3 and \S 4, we confront those
predictions with the Monte-Carlo
calculations and the simulations.

\section{The Physical Models}

To illustrate our calculations we consider two different cases 
for the linear density power spectrum. We also consider two different
physical applications: the selection function corresponding
to angular catalogs and that corresponding
to detection of weak shear.

\subsection{The power spectra}

In our first investigation we use the power spectrum
derived from the angular APM Galaxy Survey (Maddox et al. 1990), 
which has been
found to be well described by (Baugh \& Gazta\~naga, 1996) 
\begin{equation}
P_{APM}(k) \propto {k\over \left[1+(k/k_c)^2\right]^{3/2}},\label{pk}
\end{equation}
with
\begin{equation}
k_c\approx 150\ H_0/c.
\end{equation}

In the second case, we use the standard CDM model
with $\Omega_{\rm baryon}=0.05$, $H_0=50\ km/s/$Mpc
$\Omega_0=1$, $\Lambda=0$,
and an initial Harrison-Zel'dovich spectrum.
For convenience the mass fluctuation power spectrum is 
approximated by a simple analytic fit
(similar to the ones
proposed by Bond \& Efstathiou 1984),
\be
P_{CDM}(k)=A\,\disp{k\over\left(1+\left[a\,k+(b\,k)^{3/2}+
(c\,k)^2\right]^u\right)^{2/u}},
\ee
with $u=1.13$, 
$a=\disp{6.5\over 3000\ \Gamma}$,
$b=\disp{3\over 3000\ \Gamma}$
$c=\disp{1.7\over 3000\ \Gamma}$
$\Gamma=0.5$,
where the dimensional quantities have been expressed
in such a way that $c=H_0=1$.


\subsection{The angular galaxy catalogue selection function}

\begin{figure}
\centering
\centerline
{\epsfxsize=8.truecm \epsfysize=8.truecm 
\epsfbox{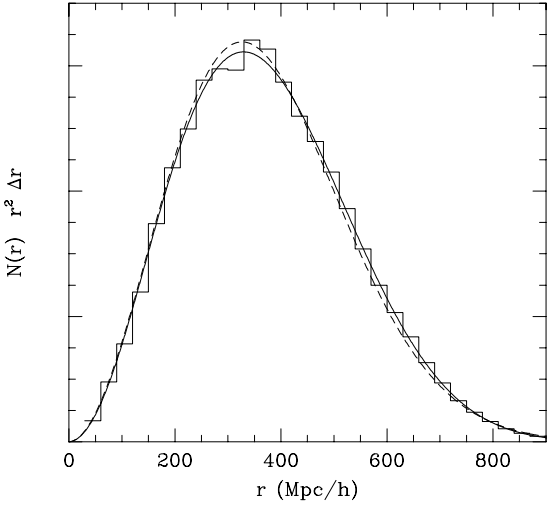}}
\caption{Comparison the theoretical (lines) and measured 
counts (histogram) in radial shells for a mock catalogues. The dashed line is
a model while the continuous line corresponds to the input APM
selection function.}
\label{nzdz}	
\end{figure}

In this case the local projected density field $\omega$ in the direction
$\gam$ is given by
\be
\delta_\gam =\int\d r\ r^2\ F(r)\ \delta(r,\gam),
\ee
where $\delta(r,\gam)$ is the local over-density
in the direction $\gam$ at the distance $r$, and $F(r)$
is the {\it normalized} selection function for the catalogue,
\be
\int\d r\ r^2\ F(r)=1.
\ee
In case of the APM angular survey the radial selection 
function is given in Gazta\~naga \& Baugh (1997, hereafter GB97). 
This function can be approximated by:
\be
F(r)\propto r^{-b}\ \exp[-r^2/D^2].
\label{selec}
\ee
We will use $b \simeq 0.1$ and $D \simeq 335 \mpc$, which provides
a good fit to the APM selection function, as shown in Figure \ref{nzdz}
(see \S\ref{nbody}, below).

We can calculate the 
moments of the distribution of the local
projected density $\delta_\theta$, smoothed at scale $\theta$
(see Bernardeau 1995, Pollo \& Juszkiewicz 1997) using
Perturbation Theory and the small angle approximation.
The second moment, or variance of angular counts-in-cells,
at the smoothing scale $\theta$, defined in (\ref{var}), 
is given by 
\be
\mg \delta^2_\theta\md=\disp{{1\over 2\pi}\ 
\int \d r\ r^4\ F^2(r)\int k\,\d k\ P(k)\ W_{2D}(k\,\theta)}
\ee
where $W$ is the window function in $k$ space. 
This variance can also
be related to an area average over the two-point angular
correlation (e.g. Gazta\~naga 1994, hereafter G94). 
Note that this expression does not rely on perturbative calculation
if the power spectrum that is used is the final one.
It uses, however the small angle approximation.
In the following we will use exclusively the top-hat window function
so that
\be
W_{2D}(k)=2 ~{J_1(k)\over k},
\ee
where $J_1$ is the spherical Bessel Function.
The third moment of smoothed angular fluctuations, defined in 
(\ref{ske}), is given by

\ba
&&\mg \delta^3_\theta\md=
{6\over (2 \pi)^2}
\int dr\ r^6 F^3(r)
\displaystyle{\left[{6\over7}\left(\int k\d k
W_{2D}^2(k\,\theta)\,P(k)\right)^2
\right.}\nonumber\\
&&+\displaystyle{{1\over 2}
\int k\,\d k\,W^2_{2D}(k\,\theta)\,P(k)\times}\\
&&\disp{\left.\int {k\ \d k\over\theta}\,
W_{2D}(k\,\theta)\,W_{2D}'(k\,\theta)\,P(k)\right]}\nonumber
\ea

So that, in case of a power law spectrum $P(k) \sim k^n$,
we have (B95),
\be
s_3^{\rm Gal.} \equiv {\mg \delta^3_\theta\md \over{\mg 
\delta^2_\theta\md^2}} =
\disp{R_3\ \left({36\over 7}-{3\over 2}\,(n+2)\right)},
\ee
with
\be
R_3=\disp{
\int r^{8-2(n+3)}\ \d r\ F^3(r)\over
\left[\int r^{5-(n+3)}\ \d r\ F^2(r)\right]^2},
\ee
for a normalized selection function.
The coefficient $R_3$ is found in practice to be of order 
unity and to be very weakly dependent
on the  adopted shape for the selection function. Note however
that the redshift evolution of the fluctuations has not been taken into 
account in this relation. This evolution should be taken into
account for catalogs having a large depth. In this case, the
geometrical factors for non-flat universes are also important
(see G94).

For the selection function given in equation (\ref{selec}) we can 
calculate $s_3^{\rm Gal.}$ explicitly,
\ba
s_3^{\rm Gal.}&=&\disp{ {8\over{3\sqrt{3}}}
\left({\sqrt{27}\over 4}\right)^b\ 
{\Gamma[{3/2-b/2}]\Gamma[3/2-n-3/2\ b]\over\Gamma[3/2-n/2-b]^2}
} \nonumber\\
&&\times\  \left({3\over 2}\right)^n
\left[{36\over7}-{3\over2}(n+2)\right].
\ea
For $b=0$ and $n=0$ we find $R_3={8\over{3\sqrt{3}}} \simeq 1.54$, while
for  $b=0$ and $n=-1$, closer to the APM case, $R_3 = {2\pi\over{3\sqrt{3}}}
\simeq 1.21$, comparable to the values given in G94.

\subsection{The weak lensing efficiency function}


In the case of weak lensing the local projected density cannot be directly
observed. We can however observe the local convergence
(see Kaiser 1995, Bernardeau et al. 1997)
$\kappa$,
\be
\kappa(\gam)=-{3\over 2}\,\Omega_0\,\int\d r\ E(r)\ 
\delta(r,\gam)
\ee
where the efficiency function, $E(r)$, depends on the
redshift distribution of the lenses. For simplicity
 we assume that the lenses
are all at redshift $z=1$ and that we live in an 
Einstein-de Sitter Universe. In this case we have
\be
E(r)={r\,(r_s-r)\over r_s}\,\left({2\over 2-r}\right)^2,
\ee
with
\be
r_s=2-\sqrt{2}.
\ee
Here the second and third moment are given by
\be
\mg \kappa^2(\theta)\md=\disp{{3\over 4\pi}\ 
\int \d r\ E^2(r)\int k\,\d k\ P(k)\ W_{2D}(k\,\theta)}
\ee
and
\ba
&&\mg \kappa^3(\theta)\md=\\
&&-{27\over 2\,(2 \pi)^2}
\int_0^{r_s}\d r\ E^3(r)
\displaystyle{\left[{6\over7}\left[\int k\d k W_{2D}^2(k\,\theta)\,P(k)\right]^2
\right.}\nonumber\\
&&+\displaystyle{{1\over 2}
\int k\,\d k\,W^2_{2D}(k\,\theta)\,P(k)\times}\nonumber\\
&&\disp{\left.\int {k\,\d k\over\theta}\,
W_{2D}(k\,\theta)\,W_{2D}'(k\,\theta)\,P(k)\right]}\nonumber
\ea

These moments can be calculated explicitly only for
particular power spectra. In the case of a power law spectrum 
of index $n$, we have (see Bernardeau et al. 1997),
\begin{eqnarray}
\displaystyle{s_3^{\rm WL}}&=\displaystyle{-\left[{36\over 7}-
{3\over 2}(n+2)\right]\,(n-1)\,(n-2)\,(n-3)^2}
\times\nonumber\\
&\displaystyle{\left[40-{2\,n\,(2\,n-1)}-{32\,n\over\sqrt{2}}
\right]/  }\label{s3z}\\
&\displaystyle{\left[16\,n\,(2\,n-5)\,(2\,n-3)\,(2\,n-1)\,
\left(2-\sqrt{2}\right)^2\right]}.\nonumber
\end{eqnarray}

In the following we will compare these P.T. results with results of
numerical simulations.

\section{A Monte-Carlo integration for $s_3$}

The aim of these Monte-Carlo integrations is to check the validity of the
small angle  approximation in PT. The integration is made here in real space
by throwing points at random in conical cells and by calculating
the averages of the second and third correlation functions.
We then identify the moments with those geometrical averages,
\ba
\mg\delta^2_\theta\md&=&\disp{1\over V_{\rm cone}^2}\,
\int_{V_{\rm cone}}d^3 \vr_1\ \d^3\vr_2\
\xi_2(\vr_1,\vr_2), \label{var}\\
\mg\delta^3_\theta\md&=&\disp{1\over V_{\rm cone}^3}\, 
\int_{V_{\rm cone}}\d^3 \vr_1\ \d^3\vr_2\ \d^3\vr_3\ 
\xi_3(\vr_1,\vr_2,\vr_3),
\label{ske}
\ea
where $\xi_2$ and $\xi_3$ are respectively the two and three-point correlation
functions in real space and $V_{\rm cone}$ is the 'volume' of the cone of angle
$\theta$ and a radial distribution given by the selection function
$F(r)$.
In Perturbation Theory they can both be expressed
in term of the linear power spectrum,
\ba
\xi_2(\vr_1,\vr_2)&=&\int\disp{\d^3\vk\over (2\pi)^3}
\ P(k)\ \exp[\ii\vk\cdot(\vr_2-\vr_1)]\\
&=&\int\disp{k^2\,\d k\over 2\pi^2}
\ P(k)\ \disp{\sin(k\,\vert \vr_2-\vr_1\vert)\over k\,\vert\vr_2-\vr_1\vert},
\nonumber
\ea
and
\ba
&&\xi_3(\vr_1,\vr_2,\vr_3)=\int\disp{\d^3\vk_1\over (2\pi)^3}
\ \disp{\d^3\vk_2\over (2\pi)^3}\\
&&\times\ P(k_1)\ \exp[\ii\vk_1\cdot(\vr_2-\vr_1)]
\ P(k_2)\ \exp[\ii\vk_2\cdot(\vr_3-\vr_1)]\nonumber\\
&&\times
\ \left[{10\over7}+{\vk_1\cdot\vk_2\over k_1^2}
+{\vk_1\cdot\vk_2\over k_2^2}
+{4\over 7}{(\vk_1\cdot\vk_2)^2\over k_1^2\ k_2^2}\right]+{\rm cyc.}
\nonumber
\ea
As shown in the appendix it is actually more convenient
to express these quantities through a set of real space
 functions.  Indeed, we can define the function $\varphi(r)$ by,
\be
\varphi(r)=\int\disp{k^2\,\d k\over 2\pi^2}\ \disp{P(k)\over k^2}\ 
\disp{\sin(k\,r)\over k\,r}
\ee
from the derivative of which the three point function can be expressed.
The Monte-Carlo integrations can then be done
in real space. They reduce to 6 dimensional integrals (see appendix)
which can be done on common work stations.

\begin{figure}
\centering
\centerline
{\epsfysize=9.0truecm 
\epsfbox{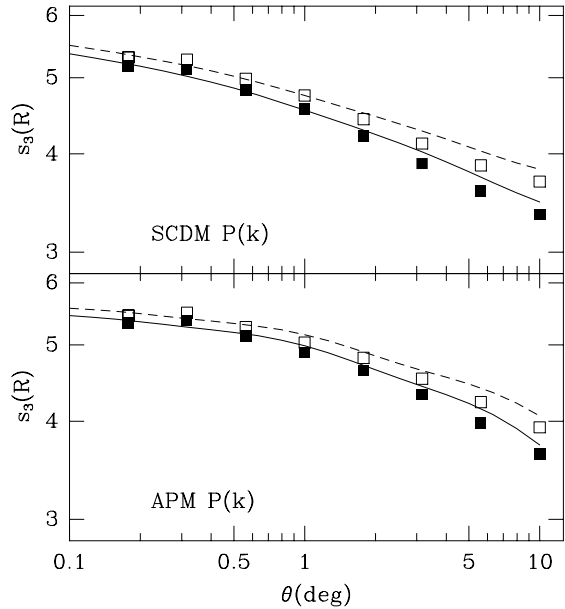}}
\caption{Angular skewness $s_3$ for galaxy catalogue selection function
from Monte-Carlo integration
(symbols) compared with small angle approximation PT predictions
(lines). The open figures and dashed line include (linear) redshift evolution.}
\label{s3ptmc}	
\end{figure}

\begin{figure}
\centering
\centerline
{\epsfysize=9.0truecm 
\epsfbox{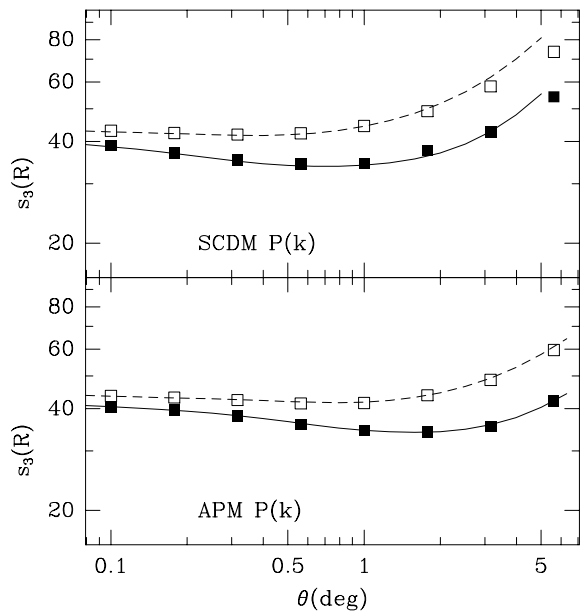}}
\caption{Angular skewness $s_3$ for weak lensing efficiency function
from Monte-Carlo integration
(symbols) compared with small angle approximation PT predictions
(lines). The opened figures and dashed line include (linear) 
redshift evolution.}
\label{s3lptmc}	
\end{figure}

In Figures \ref{s3ptmc} and  \ref{s3lptmc}, we compare
the results of the Monte-Carlo integration (symbols)
for  $s_3$ 
with the small angle approximation (curves) in PT.
We show both the case with (linear) redshift evolution
(open figures and dashed line), which would correspond to
real observations,
and also the case without redshift evolution, 
as corresponds to the comoving output from N-body
simulations (see below). 

There is an excellent agreement in the comparison
 specially in the weak lensing case.
For the galaxy selection function there are some small discrepancies
at the largest scales, but they are  smaller than the typical errors
in the estimation from the observations.
 Thus the small angle approximation seems
to be quite good for most applications up to 10 degree scales.

\section{The numerical $N$-body results}
\label{nbody}

We study two shapes of the power spectrum: the APM model and the CDM
model, as described in Section \S2.
We use several sets of N-body simulations with parameters
 shown in Table
\ref{tab:param}. The letters refer to different realizations of
each model: 5 for APM1, 2 for APM2 and 5 for CDM.
 The APM like models are described
in detailed in GB97, while the CDM simulations are from
GB95, and correspond to the standard model: $\Gamma=0.5$.

\begin{table}
\begin{center}
\caption[dummy]{Simulation parameters}
\label{tab:param}
\begin{tabular}{lcccc}
\hline
\hline  
run       & number    & mesh    & $L_{box} $   \\ 
            & of particles         &         & $(\mpc) $  \\ \hline
\\
APM1(a)-(e)     & $126^{3}$ & $128^{3}$ & 400   \\
APM2(a)-(b)     & $200^{3}$ & $128^{3}$ & 600         \\ 
CDM(a)-(e)     & $126^{3}$ & $128^{3}$ & 378         \\  \\
\hline
\label{tab:prop}
\end{tabular}
\end{center}
\end{table}

For the APM like models we use the output time corresponding to the
measured APM amplitude, $\sigma_8 \simeq 0.8$, before it is scaled
up to account for clustering evolution (see  Gazta\~{n}aga 1995).
For CDM models we use $\sigma_8=1$, unless otherwise stated.
As we are using outputs in comoving coordinates, there is
no redshift evolution within one output. This is not the case
in the real Universe, but we take this into accunt 
in the theoretical predictions in a straightforward way (G94,
see Figures \ref{s3ptmc} and  \ref{s3lptmc}).

\begin{figure*}
\centering
\leftline
{\epsfysize=9.0truecm 
\epsfbox{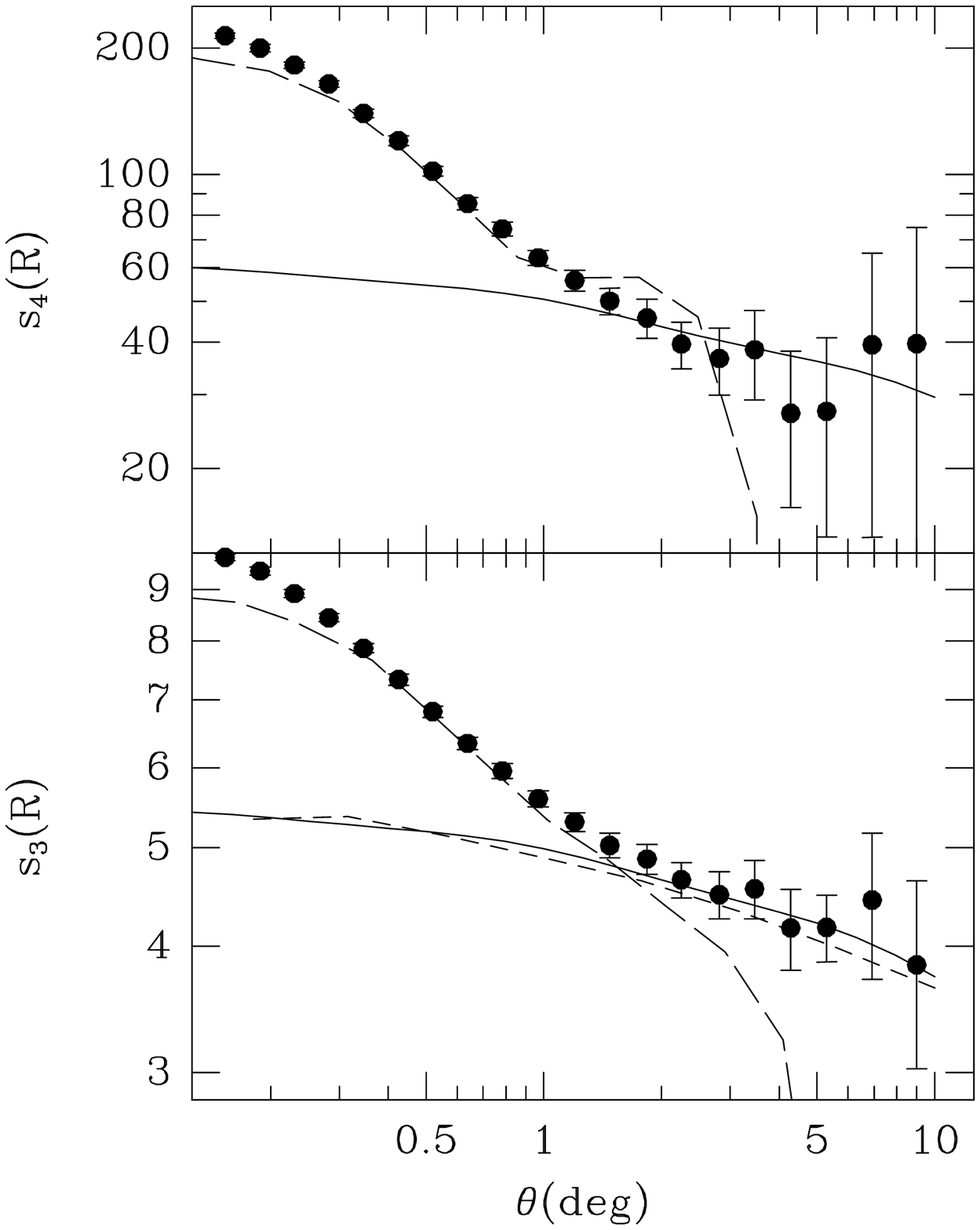}}
\vskip -9 truecm
\rightline
{\epsfysize=9.0truecm 
\epsfbox{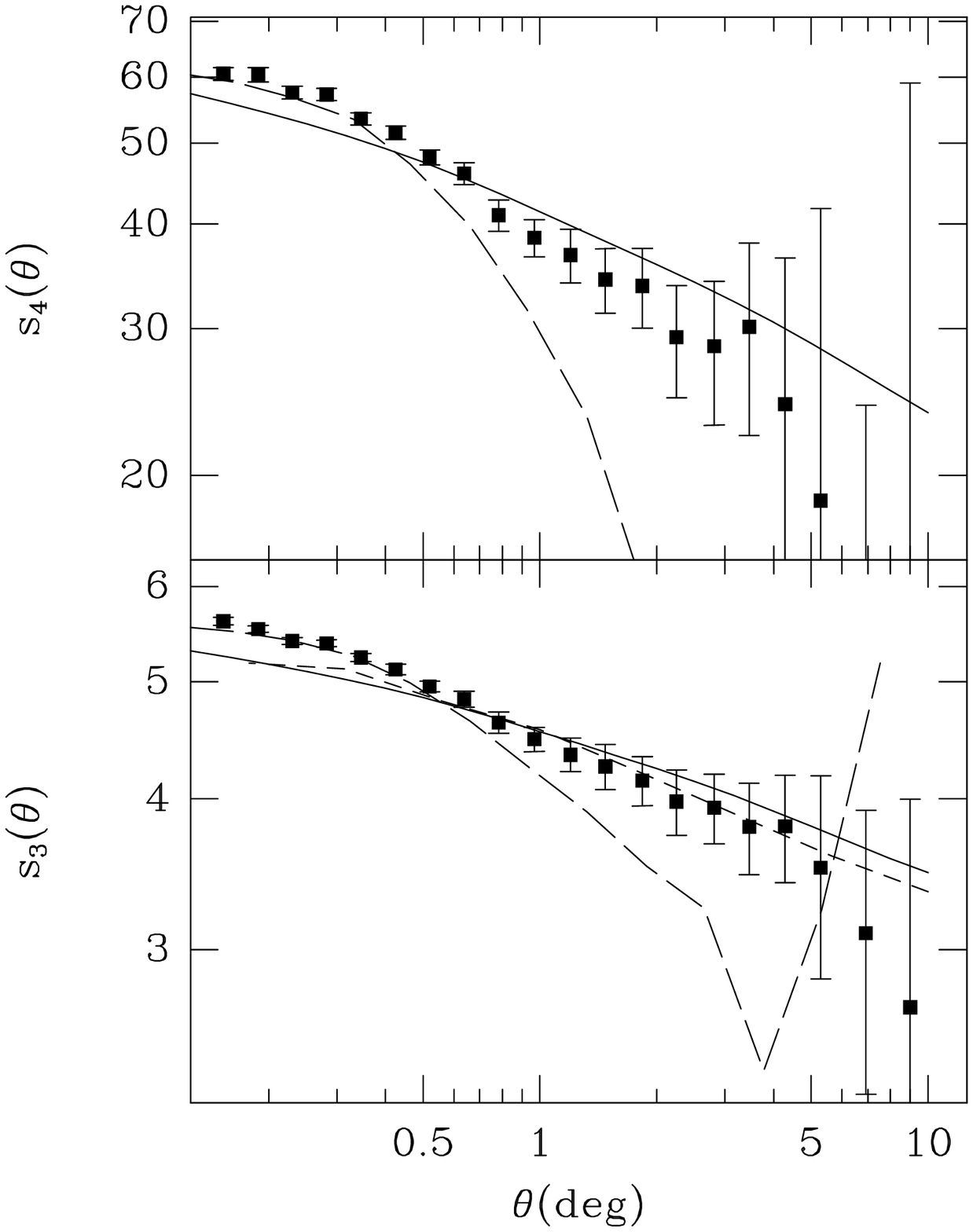}}
\caption{Angular skewness $s_3$ (bottom) and kurtosis $s_4$ (top)
 in mock maps with the APM power spectrum (left panels) and the
CDM power spectrum (right panels) for the galaxy catalogue selection function.
Results are symbols with errorbars and are
compared with small angle approximation PT predictions
(continuous line). The short-dashed
line shows the Monte Carlo
integration. The long-dashed line are results from smaller mock
catalogues with the (APM) observational mask.}
\label{s34c600x}	
\end{figure*}


To produce the angular catalogues we first 
select an arbitrary point in the simulated box to be the local 
'observer'. We include a simulated particle at comoving
coordinate $r$ from the observer
with probability given by the selection function $F(r)$, in the case
of galaxy clustering, 
or the efficiency function in the case of weak lensing, as
described in Section \S2. 
As the simulation is done in a periodic box, we replicate the box to cover
the total radial extent of the APM (over $1800 \mpc$). 
 The main difference with GB97,
is that they apply the APM angular survey mask,
 including plate shapes and holes, whereas here we use full sky maps,
which cover a larger area and have no boundaries.
In our full sky mock catalogues there could be a certain degree of 
repetition,
but the errors are estimated form the dispersion in different catalogues.
By comparing the results from different box sizes we have verified that this
replication of the box does not introduce any spurious effects.
The total number of particles in the mock angular catalogues
 is about $9 \times 10^6$ for the APM
galaxy selection function and about  $25 \times 10^6$
for the weak lensing case. 

Figure \ref{nzdz} shows a comparison between the expected 
number of galaxies, $N(r) r^2 \Delta r$, at different radial
depths (in comoving coordinates $r$) given by the 
input selection function, as compared to the measured counts
in a mock catalogue. The dashed line is
the model in equation (\ref{selec}) for  
$b \simeq 0.1$ and $D \simeq 335 \mpc$,
 while the continuous line is the input APM
selection function (GB97).

From each realization we produce 
several mock catalogues by choosing different positions for the observer. 
Because of the selection function, catalogues from different observers
are not necessarily correlated. Three different observers are used
for the smaller boxes. For the larger boxes the corresponding number is
10 observers. For these numbers, results from different observers do not seem
to be correlated. Thus the total number of mock catalogues 
is 15 for APM1 and CDM, and 20 for APM2.

Figures \ref{s34c600x} shows the results for
$s_3$ and $s_4$ for the galaxy CDM and APM2
mock catalogues. 
Results for APM1 catalogues, which are not shown in the Figures,
 agree well with the ones in APM2, indicating that the size of the
simulation  box is large enough.

There is very good agreement between the N-body results and
 PT theory at scales
 $\theta \simgt 2 \, deg$ for the APM model. For CDM, there is agreement 
within the errors for
$\theta \simgt 1 \, deg$, although this is not as good as for the APM model.
This could be due to finite volume effects, which seem more
important for CDM (see below); note also that the CDM errors are larger and
that the CDM maps come from a simulation with a smaller volume.
Scales $\theta \simeq 1 \, deg$ in the CDM models could also be affected
by projection effects that tend to overcompensate non-linearities
(see below); this is more important for CDM which has a higher
normalization ($\sigma_8=1$).

\begin{figure}
\centering
\centerline
{\epsfysize=9.0truecm 
\epsfbox{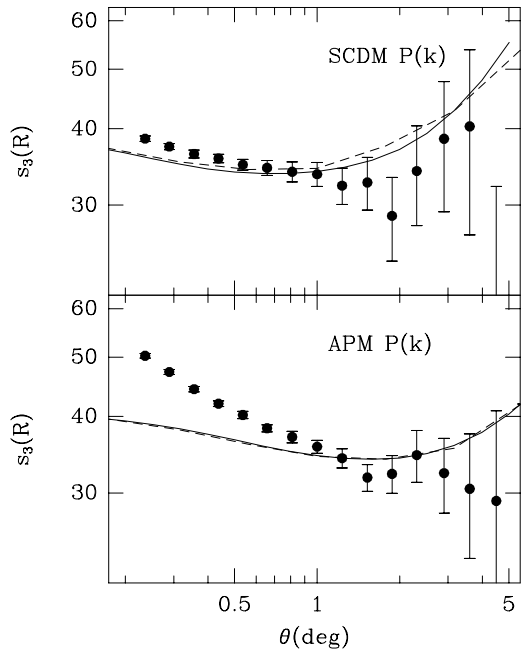}}
\caption{Angular skewness $s_3$ for weak lensing efficiency function 
from simulations
compared with small angle approximation PT predictions
(continuous line). The short-dashed line shows the Monte Carlo
integration.}
\label{s3lens}	
\end{figure}

Figure \ref{s3lens} shows a comparison of PT with
 the skewness estimated in the mock maps made
with the weak lensing efficiency function. The errors are larger here
than in the galaxy case
because the depth (and volume) is larger and therefore a given
physical scale corresponds to a smaller angular scale.

\begin{figure}
\centering
\centerline
{\epsfxsize=8.truecm \epsfysize=8.truecm 
\epsfbox{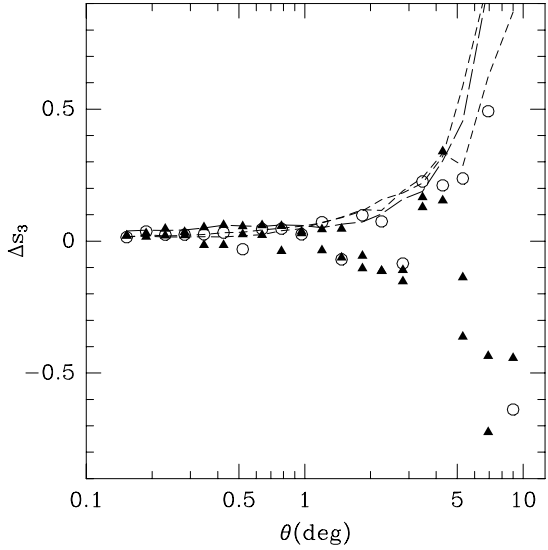}}
\caption{Relative 
errors in the the distribution of values of $s3$ from different
maps as a function of the angular scale. Dashed line corresponds to the
$rms$ value:  $ \mg(\Delta s_3)^2\md^{1/2}$, whereas the points
correspond to a measure of the skewness: $ \mg(\Delta s_3)^3\md^{1/3}$.
Open circles and long dashed lines correspond to the maps from the
largest box simulations. Filled triangles and short-dashed lines
show resulsts from the small boxes.}
\label{difs3}	
\end{figure}

\subsection{Sampling variance}

We find that it is crucial
to use a large number of catalogues in order to have a robust 
estimation
at the largest angular scales. This is illustrated in 
Figure \ref{difs3}, which shows moments of the distribution of 
relative errors  in $s_3$, i.e. $\Delta s_3 \equiv {s_3/\bar{s_3}}-1.$,
from map to map around the mean value, $\bar{s_3}$. The $rms$ error
(dashed line) increases with angular scale as expected. The
skewness of the error distribution,  $ \mg(\Delta s_3)^3\md^{1/3}$ (points
in the figure), has large 
fluctuations, but seems to have a tendency to go negative 
on the largest scales in each catalogue.
We show results for both the large APM simulations
(open circles and long dashed lines) and both CDM and APM $L=400 \mpc$
simulations. 
This result indicates that it is more likely to find smaller values
of $s_3$, when doing a smaller sampling, e.g. in a single map.

The above arguments are also illustrated by comparing the results
from the full sky mock catalogues with the mean values 
using 10 mock catalogues with the APM mask 
(e.g. GB97). The latter not only
cover a smaller fraction of the sky (only about $10\%$) 
but are also subject to boundary effects. The results for these
{\it realistic} maps are shown as long-dashed lines in Figures
\ref{s34c600x}.
 The finite volume and boundary effects
seem important on scales $\theta \simgt 1 \, deg$ for CDM and about
 $\theta \simgt 2 \, deg$ for APM.

\subsection{Time evolution}

Perturbation theory predicts time independent values of the
skewness, in the limit of small variance. For large variance,
N-body simulations find that
the 3D skewness, $S_3$, increases with time due to non-linear
effects (e.g. Baugh, Gazta\~naga \& Efstathiou 1994,
Colombi et al. 1997).  Non-linearities also erase
the shape dependence in the hierarchical structure, e.g. the bispectrum,
(see Scoccimarro et al. 1997)
reproducing a simple tree-level hierarchy, independent of the
shape of the configurations. When this happens 
there is a simple relation between the angular and 3D skewness:
$s_3 \simeq R_3 \, S_3$ (see G94), so that the corresponding
2D projected amplitudes are closer to the 3D PT results.
Thus, on quasi-linear scales,
the relation $s_3 \simeq R_3 \, S_3$
typically underestimates the projection  effects, while it should
be more accurate on smaller scales, were PT results will 
overpredict the projection effect.

\begin{figure}
\centering
\centerline
{\epsfysize=9.0truecm 
\epsfbox{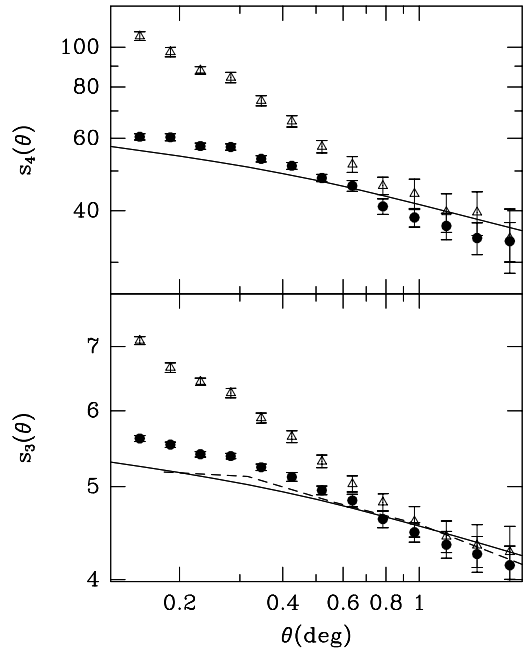}}
\caption{Angular skewness (bottom) and kurtosis (top) in 
CDM mock maps from 
different output times: $\sigma_8=1$ (filled circles)
and  $\sigma_8=0.5$ (open triangles). Continuous lines show the 
small angle approximation PT predictions. The short-dashed
line shows the Monte Carlo
integration.}
\label{s34c189}	
\end{figure}

Figure \ref{s34c189} shows a comparison of $s_3$ and $s_4$
from mock maps drawn from two different outputs in the CDM
galaxy model. Contrary to what one would expect, the later output
seems closer to PT results. This is because of the projection
effects mentioned above, which compete with the non-linear growth in $s_3$.
In fact,  the slight disagreement at the smallest
angular scales between the results of
the full sky map (symbols with errorbars) and the one with the APM mask
(long-dashed lines) shown in Figure \ref{s34c600x}
is also due to the fact that we have used slightly 
different output times in each case.

\section{Conclusion}

We conclude from this analysis that the P.T. results for
the skewness and kurtosis are accurate
at  degree scales or larger for the projected density.
Comparisons between Monte-Carlo integrations and results
obtained from the small-angle approximation are in good agreement.
A departure starts to be noticeable only for smoothing
scales above 5 degrees, but even there it is small compared with
typical errors.

The $N$-body results allow us to estimate the small angular scales
at which  nonlinear effects start to affect the values of $s_3$.
We found these effects to be important below $0.5 \, deg$
for the lensing case, and below $1 \, deg$ for the galaxy selection function,
with a rapid growth of $s_3$ at small scale. This however depends
strongly on the initial power spectrum. This effect is found to be
more important for the measured APM power spectrum
than for the CDM one.

We have found two effects that might prove important in comparing
perturbation theory with observations of angular clustering, and
in particular for the APM. First,
volume and boundary effects are important on scales
$\simgt 2$ deg and tend to produce smaller values of $s_3$ 
and $s_4$. We argue that  this is not because the mean is biased, but because
the distribution of errors seems to be negatively skewed
(e.g. Figure \ref{difs3}) and it is therefore more probable to find smaller
values. Second, the tree-level hierarchy
model for projections commonly used in the literature
(e.g. by Groth \& Pebles 1977, Fry \& Peebles
1978,  Szapudi, Szalay \& Boschan 1992, Szapudi et al. 1996), 
and in particular for the APM (G94) is not accurate on 
quasi-linear scales, as indicated in B95, 
because it underestimates the projection  factors.
At small scales, non-linear effects tend 
to mix these two cases (e.g. Figure \ref{s34c189}), while at
larger scales this competes with large volume effects.
The errors involved are of the order of $20\%$ in $s_3$,
but  more quantitative analysis of this point
and implications for the APM observations
will be presented elsewhere.

To have a better insight into these nonlinear effects, and how it
depends on the shape of the power spectrum, it could be fruitful
to extend the recent results of Scoccimarro (1997) on the
one-loop correction of the bispectrum to the projected third moment.

\section*{Acknowledgments}

We would like to thank Fermilab group for their kind hospitality 
during our visit in  the fall 1996. EG would like to thank Carlton Baugh and
Josh Frieman for helps and comments on the manuscript.
E.G. acknowledges support from CIRIT (Generalitat de
Catalunya) grant 1996BEAI300192, CSIC, DGICYT (Spain), project
PB93-0035, and CIRIT, grant GR94-8001.

\section*{Appendix: The expression of the three-point correlation 
function in real space} 

We want to express the three point correlation function provided by
Perturbation Theory (Peebles 1980, Fry 1984),
\ba
&&\xi_3(\vr_1,\vr_2,\vr_3)=\int\disp{\d^3\vk_1\over (2\pi)^3}
\ \disp{\d^3\vk_1\over (2\pi)^3}\\
&&\times\ P(k_1)\ \exp[\ii\vk_1\cdot(\vr_2-\vr_1)]
\ P(k_2)\ \exp[\ii\vk_2\cdot(\vr_3-\vr_1)]\nonumber\\
&&\times
\ \left[{10\over7}+{\vk_1\cdot\vk_2\over k_1^2}
+{\vk_1\cdot\vk_2\over k_2^2}
+{4\over 7}{(\vk_1\cdot\vk_2)^2\over k_1^2\ k_2^2}\right]+{\rm cyc.}
\nonumber
\ea
in real space only. The reason is that the integration
in $k$ space with real space top-hat window function
is hardly possible because the integrals converge very slowly.

Let us introduce the function $\varphi(r)$,
\ba
\varphi(r)&=&\int\disp{\d^3 \vk\over (2\pi)^3}\ \disp{P(k)\over k^2}\ 
\exp(\ii\vk\cdot\vx)\\
&=&
\int\disp{k^2\,\d k\over 2\pi^2}\ \disp{P(k)\over k^2}\ 
\disp{\sin(k\,r)\over k\,r}.
\ea
Then we can notice that
\ba
\grad\xi(\vx)&=&-\ii\int{\d^3\vk\over(2\pi)^3}\,\vk\,\exp(\ii\vk\cdot\vx)\\
&=&\disp{\vx\over x}\,\disp{\d\xi(x)\over \d x}
\ea
(and a similar property for $\varphi$) so that
\ba
&&\grad\xi(\vx_{12})\cdot\grad\varphi(\vx_{13})\\
&&=-\int\disp{\d^3\vk_1\over (2\pi)^3}\,\disp{\d^3\vk_2\over (2\pi)^3}\,
\disp{\vk_1\cdot\vk_2\over k_2^2}
\ \exp(\ii\vk_1\cdot\vx_{12}+\ii\vk_2\vx_{13})\nonumber\\
&&=\disp{\vx_12\cdot\vx_{13}\over x_{12} x_{13}}\ 
\disp{\d\xi(x_{12})\over \d x_{12}}\ \disp{\d\varphi(x_{13})\over \d x_{13}}
\ea
and one recognizes one term that intervenes in the expression
of the three-point function.

To complete the calculation one can also notice that,
\ba
&&\varphi_{ij}(\vx_{12})\varphi_{ij}(\vx_{13})\\
&&=\int\disp{\d^3\vk_1\over (2\pi)^3}\,\disp{\d^3\vk_2\over (2\pi)^3}\,
\disp{(\vk_1\cdot\vk_2)^2\over k_1^2\,k_2^2}
\ \exp(\ii\vk_1\cdot\vx_{12}+\ii\vk_2\vx_{13})\nonumber\\
&&=
\disp{\varphi'(x_{12})\,\varphi'(x_{12})\over x_{12}\,x_{13}}
+\disp{\varphi''(x_{12})\,\varphi''(x_{13})}+\nonumber\\
&&
\disp{(\vx_{12}\cdot\vx_{13})^2\over x_{12}\,x_{13}}\,
\left(\varphi''(x_{12})-{\varphi'(x_{12})\over x_{12}}\right)
\,\left(\varphi''(x_{13})-{\varphi'(x_{13})\over x_{13}}\right).\nonumber
\ea
Finally remarking that
\be
\varphi''(x)=-\xi_2(x)-{2\over x}\varphi'(x)
\ee
one can express the three point correlation function in real space
with the function $\xi_2(x)$, $\xi'_2(x)$ and $\varphi'(x)$ only:
\ba
&&\xi_3(\vr_1,\vr_2,\vr_3)=\disp{10\over 7}\xi_2(x_{12})\,\xi_2(x_{13})+
\nonumber\\
&&\disp{4\over 7}\disp{\varphi'(x_{12})\over x_{12}}\,
\disp{\varphi'(x_{13})\over x_{13}}+\nonumber\\
&&\disp{4\over 7}\,
\left(\xi_2(x_{12})+2\disp{\varphi'(x_{12})\over x_{12}}\right)\,
\left(\xi_2(x_{13})+2\disp{\varphi'(x_{13})\over x_{13}}\right)\,\nonumber\\
&&-u\,\left[\xi_2'(x_{12})\,\varphi'(x_{13})+
\xi_2'(x_{13})\,\varphi'(x_{12})\right]+
\\ \nonumber
&&\disp{4\over 7}\,u^2\,
\left(\xi_2(x_{12})+3\disp{\varphi'(x_{12})\over x_{12}}\right)\,
\left(\xi_2(x_{13})+3\disp{\varphi'(x_{13})\over x_{13}}\right)+{\rm cyc.},
\ea
where
\be
u\equiv \disp{\vx_{12}\cdot\vx_{13}\over x_{12}\,x_{13}}.
\ee
This result generalizes the one obtained by Fry (1984) for power law
spectra.

The Monte-Carlo computations of the geometrical averages of
$\xi_2$ and $\xi_3$ can then be done
in real space, provided the functions $\xi_2$, $\xi_2'$
and $\varphi'$ are known. Then the integrations can be
reduced to a 6-dimensional integrals by direct integration over the azimuthal
angles (see fig. 3).

\begin{figure}
\vspace{10 cm}
\special{hscale=60 vscale=60 voffset=10 hoffset=40 psfile=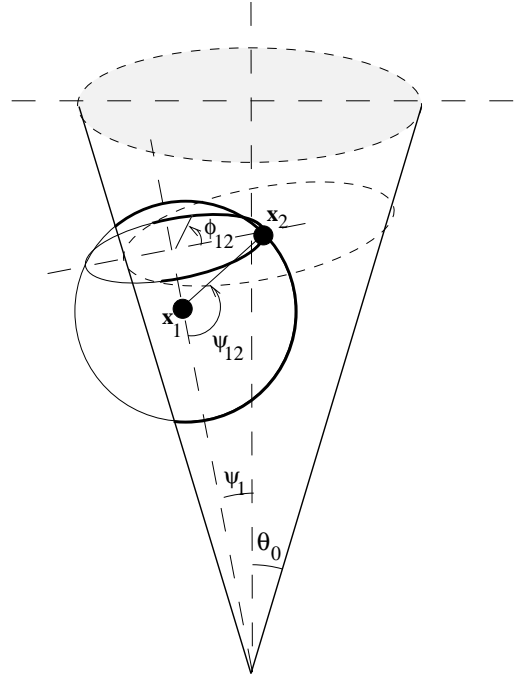}
\caption{Description of the variables used in the Monte
Carlo Integration. The distance of $\vx_1$ to the origin, the
distance between $\vx_1$ and $\vx_2$, the angle $\Psi_1$ and
the angle $\Psi_{12}$ are all chosen randomly. 
The integration over $\phi_{12}$ can be done explicitly since the distances
of $\vx_2$ to the origin and to $\vx_1$ remain then constant.
}
\end{figure}

\end{document}